\documentclass{sig-alternate}

\begin{document}
%
\permission{Copyright is held by authors/owners.}
\conferenceinfo{ECML/PKDD Discovery Challenge,}{September
20th,2010.} \copyrightetc{Barcelona, Spain}
\CopyrightYear{2010} 

\title{Evaluating Web Content Quality via Multi-scale Features}
%
%
%
%
%

\numberofauthors{4} 
%
\author{
%
%
\alignauthor
Guang-Gang Geng\\
       \affaddr{CNNIC, Computer Network Information Center}\\
       \affaddr{Chinese Academy of Science}\\
       \affaddr{Beijing 100190, P. R. China}\\
       \email{guanggang.geng@gmail.com}
\alignauthor
Xiao-Bo Jin\\
       \affaddr{School of Information Science and Engineering, Henan University of Technology}\\
       \affaddr{Henan 450001, P. R. China}\\
       \email{xbjin@nlpr.ia.ac.cn}
\and
\alignauthor Xin-Chang Zhang\\
       \affaddr{Shandong Computer Science Center, Shandong Academy of Sciences}\\
       \affaddr{Shandong 250014, P. R. China}\\
       \email{xinczhang@hotmail.com}
 \alignauthor
De-Xian Zhang\\
       \affaddr{School of Information Science and Engineering, Henan University of Technology}\\
       \affaddr{Henan 450001, P. R. China}\\
       \email{zdxzzit@hotmail.com}}


\date{30 July 1999}

\maketitle
\begin{abstract}
Web content quality measurement is crucial to various web content
processing applications. This paper will explore multi-scale
features which may affect the quality of a host, and develop
automatic statistical methods to evaluate the Web content quality.
The extracted properties include statistical content features, page
and host level link features and TFIDF features. The experiments on
ECML/PKDD 2010 Discovery Challenge data set show that the algorithm
is effective and feasible for the quality tasks of multiple
languages, and the multi-scale features have different
identification ability and provide good complement to each other for
most tasks.

\end{abstract}




\section{Introduction}
The evaluation of Web content quality plays an important role for
various Web content processing applications, such as search engine,
Web archiving service and Internet directory, etc; but how to
evaluate the quality of the Web content?
In the past, most data quality measures were developed on an ad hoc
basis to solve specific problems, and fundamental principles
necessary for developing stable metrics in practice were
insufficient \cite{dataQuality2010}. In the research of Web content
quality assessment, computational models that can automatically
predict the Web content quality should be focused on.


Web spam can significantly deteriorate the quality of search engine
results, but high quality is more than just the opposite of Web
spam. ECML/PKDD 2010 Discovery Challenge (DC2010) aims at more
aspects of the Web sites. DC2010 wants to develop site-level
classification for the genre of the Web sites (editorial, news,
commercial, educational, ``deep Web" or Web spam and more) as well
as their readability, authoritativeness, trustworthiness and
neutrality \cite{dc2010}.

Statistical learning methods have demonstrated their effectiveness
for many classification problems, such as Web spam detection, text
categorization and anti-phishing \cite{CC:Know} \cite{www2009}
 \cite{tfidf} \cite{wsc2008} \cite{phishing} \cite{url}, which inspires us to evaluate Web
content quality with statistical learning algorithms. In this paper,
we will explore a series of features from multiple views, and
compare their effectiveness for Web content quality assessment.

The rest of this paper is organized as follows. Section 2 first
introduces the multi-scale features extraction, among which the
description of TFIDF and host level link features extracting are our
focus. Then it discusses the feature fusion strategy. Section 3
gives the Web content quality assessment method. Section 4 presents
our experiment results.
Finally, section 5 draws the conclusion and discusses the future
work on Web content quality assessment.

\section{Features Extraction}
In this section, we will describe multi-scale features extracted
from four different views, including content statistics features,
page level link related features, host level link related features
and text features(TFIDF) \cite{tfidf}, and give the feature fusion
strategy.

\subsection{Content, Link and TFIDF Features}
The content features and page level link features used here are
provided by the ECML/PKDD 2010 Discovery Challenge organization
committee, i.e. content-based features and link-based features
\cite{dc2010}.


We compute the TFIDF\cite{tfidf} features with term frequencies and
document frequencies provided by DC2010\cite{dc2010}:
\begin{equation}\label{tfidf}
a_{ik} = f_{ik}\times{\log{\frac{N}{n_i}}}
\end{equation}
where $a_{ik}$ is the weight of word $i$ in document $k$, $f_{ik}$
 the frequency of word $i$ in document $k$, N the number of
documents in the collection, and $n_i$ the total number of the word $i$ occurs
in the whole collection.

After computing $a_{ik}$, feature selection is performed. Feature
selection attempts to remove non-informative terms in order to
improve the classification performance and reduce the computation
complexity. In this paper, we select information gain (IG) for
feature selection. IG measures the number of bits of information
obtained for the category prediction by knowing the presence or
absence of a word in the document. Information gain has been proved
to be one of the most effective feature selection methods for text
categorization\cite{tfidf}, statistical spam filtering\cite{Stat04}
and  information retrieval\cite{ir}, etc.

\subsection{Host Level Link Related Features}\label{feaHost}
PageRank\cite{pr} is one of the most famous link analysis
algorithms, which reflects the importance of Web pages. With the
growing prevalence of link spam, PageRank scores become unreliable
as a quality measure. Considering the hypotheses which benign nodes
tend to link to other high quality nodes and malicious nodes are
mainly linked by low quality nodes, we will extract a series of host
link analysis features and attempt to mine the quality relations
from the topology dependency.

Let $weight=f(n)$ be a weighting function, where $n$ is the number
of links between any host pair $(h,v)(h \in V,v \in V)$ and $E$ be
the set of edges with $weight\geq{W}$, then the host graph $G$ can
be defined as $G = (V,E,weight)$. Considering the topological
dependencies of low and high quality nodes, the following features
related to host graph can be extracted:

\begin{eqnarray}
F_1(h) & = & Measure(h) \\
F_2(h) & =
&\frac{\sum_{v\in{Inlink(h)}}Measure(v)*weight(h,v)}{\sum_{v\in{Inlink(h)}}{weight(h,v)}}\\
F_3(h) & = &
\frac{\sum_{v\in{Outlink(h)}}Measure(v)*weight(h,v)}{\sum_{v\in{Outlink(h)}}{weight(h,v)}}
\end{eqnarray}

where $Measure\in\{$HostRank(Host Level PageRank), DomainPR,
Truncated PageRank($T=1,2,\cdots$), Adaptive Estimation of
Supporters($d=1,2,\cdots$)$\}$\footnote{In practice, in order to
make things easier, we use the logarithm of all these
values.}\cite{trun06}, HostRank is computed based on the host graph
$G$, and DomainPR is the rank value of a host corresponded domain,
which is queried from http://toolbarqueries.google.com, for example,
the DomainPR of impressum.dukemaster.eu is the same as that of
dukemaster.eu. $\{h,v\} \subseteq V$,
$Inlink(h)$ is the inlink set of $h$ and $Outlink(h)$ is the outlink
set of $h$. $weight(h,v)$ is the weight of host $h$ and $v$,
$weight(h,v)\in \{1,\log(n),n\}$, where $n$ is the number of
hyperlinks between $h$ and $v$.

In our experiments, $W=1$, $T \in \{1,2,3,4\}$, $d \in \{1,2,3,4\}$
and $weight(h,v)\in \{1,n\}$. Finally, we extract 50 host level link
features.

\subsection{Feature Fusion Strategy}

\begin{figure}[!htp]\centering
\includegraphics[angle=0, width=0.48\textwidth]{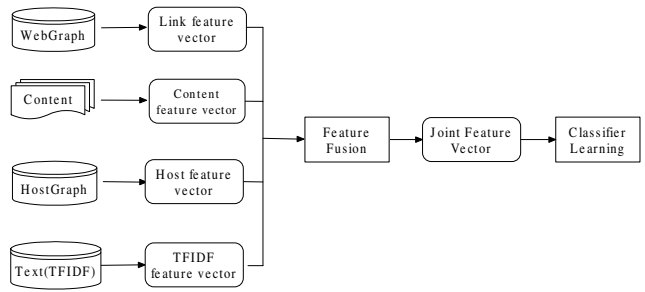}
\caption{ Flow Chart of Web Content Quality Assessment via
Multi-scale Features.}\label{fusion}
\end{figure}

To analyze the effectiveness of features of different scales, we
train classifiers with the fusion of different features.
Fig.\ref{fusion} shows the flow chart of fusion strategy with all
the above-mentioned features.

\section{Web Content Quality Assessment} \label{qas}
\subsection{Quality Assessment Strategy}
Web content quality assessment is in fact a quality prediction
problem. In ECML/PKDD 2010 Discovery Challenge, the quality value is
defined based on genre, trust, factuality and bias. Typically,
DC2010 gives the discrete value empirically:
The Spam host has quality 0; News/Editorial and Educational sites
are worth 5; Discussion hosts are worth 4 while others are worth 3.
DC2010 also gives 2 bonus scores for Facts or Trust, but subtracts 2
for Bias hosts.

In general, we can first classify the Web sites according to the
categories: Web Spam, News/Editorial, Commercial,
Educational/Research, Discussion, Personal/Leisure, Neutrality, Bias
and Trustiness. Then we further compute the Web content quality with
the criteria given by DC2010. However, the state of art
classification methods may be inappropriate for the ranking:
\begin{itemize}
  \item Most given classes are imbalanced. Therefore, training
  an effective model is difficult.
  \item The predicted probability for every class cannot be
  fully used to rank the Web content quality.
  \item The discrete predictions are unfavorable to ranking the hosts.
\end{itemize}

Considering the Web content quality values are discrete values, we
first treat the Web content quality assessment as a multi-class
classification problem, thus the aforementioned shortcomings will be
overcome in great degree. Then based on the predicted probabilities
of samples belonging to each classes, the quality values of the Web
sites are computed as follows:

\begin{equation}\label{quality}
Quality(h)=\sum_{i=0}^{N-1}{P_i(h)\times{Q(i)}}
\end {equation}
where $N$ is the number of classes, $Quality(h)$ is the quality of
host $h$, $P_i(h)$ is the predicted probability that the host $h$
belongs to class $i$, $\sum_{i=0}^{N-1}{P_i(h)} = 1$. $Q(i)$ is the
quality value of class $i$, for ECML/PKDD 2010 Discovery Challenge
quality tasks (Task2 and Task3), $Q(i)=i$, $N=10$, i.e. $i \in
\{0,1,\cdots,9\}$.


\subsection{Learning Algorithms}

For the above-mentioned Web content quality assessment strategy, the
most important is to predict the posterior probability of examples
belonging to each class effectively. Then, how to estimate the
posterior probability as accurate as possible? Fan et al. \cite{pos}
argue that randomized decision tree methods effectively approximate
the true probability distribution using the decision tree hypothesis
space. When bagging\cite{bagging} is applied to C4.5\cite{c45}, each
random tree is computed based on a bootstrap of the training
samples, which further optimizes the posterior probability
predictions.

\section{Experiments}


\subsection{Data Collection} We realize our algorithms on
ECML/PKDD 2010 Discovery Challenge dataset \cite{dc2010}.
In the experiments, we use all the labeled samples of English host
as the training samples set for Task1 and Task2. DC2010 only
provides few labeled samples for French and German Tasks. We put all
the labeled examples including English, French and German into
training set for the multilingual quality tasks(Task3). The test set
which we use in this paper is the test set for DC2000 contest.


In our experiments, we assume that the host with www and the
www-less version have the same quality. After removing the
duplicated samples, we obtain the English training set with 2114
samples, French training set with 2400 samples, and German training
set with 2400 samples.

\subsection{Features} We use all the content-based features and
link-based features provided by DC2010\cite{dc2010}. For TFIDF
features, we select 500 dimensions with the top information gain
values. We have also done experiments by selecting 1000, 1500 and
2000 TFIDF features and find there is no obvious difference for the
performance. Besides, we extract 50 host level link features as
mentioned in section \ref{feaHost}.

\subsection{Learning Algorithm and Evaluation}

As described in section \ref{qas}, the machine learning algorithm we
use is bagging, with C4.5 decision tree as the weak classifier. In
the experiments, the iterations of C4.5 in bagging are 90.

Normalized Discounted Cumulative Gain(NDCG)\cite{ndcg} is a measure
of effectiveness of a Web search engine algorithm or related
applications, which is often used in information retrieval. NDCG is
also employed for evaluating the submissions for ECML/PKDD 2010
Discovery Challenge \cite{dc2010}.
As for the detailed evaluation, please refer to DC2010 evaluation
\cite{rule}.



\subsection{Experiment Results}

Table \ref{task1} describes the NDCG performance with different
features on Discovery Challenge 2010 Task1. In line 1, L denotes the
page level link related features; H denotes the host level link
features; C denotes content statistical features; T denotes the
TFIDF features, HCT denotes the fusion of host level link features,
content features and TFIDF features; and LHCT denotes the fusion of
all the above-mentioned different scale features. The first column
of the tables shows the subtask in Task1. The column 2 to 7 are the
performances of the quality assessment method with different
features on all the subtasks.

\begin{table}[!htp]\label{task1}
 \caption{Comparisons of Web content quality assessment performance with
different features on Task1(NDCG)}
 \centering
\begin{tabular}[!htp]{p{1.5cm}p{0.7cm}p{0.7cm}p{0.7cm}p{0.7cm}p{0.7cm}p{0.7cm}}
\hline Task&L&H&C&T&HCT&LHCT\\
\hline \hline Spam&0.628&0.789&0.784&0.756&\textbf{0.830}&0.807\\
News&0.549&0.589&0.625&0.743&0.740&\textbf{0.748}\\
Commercial&0.715&0.741&0.753&0.88&\textbf{0.883}&\textbf{0.883}\\
Educational&0.726&0.808&0.805&0.872&\textbf{0.885}&0.884\\
Discussion&0.638&0.573&0.768&0.822&0.784&\textbf{0.79}\\
Personal&0.594&0.728&0.768&0.804&\textbf{0.828}&0.827\\
Neutrality&0.605&0.511&0.426&0.438&0.465&\textbf{0.495}\\
Bias&0.525&\textbf{0.606}&0.518&0.525&0.51&0.549\\
Trustiness&\textbf{0.526}&0.506&0.472&0.358&0.485&0.441\\
\hline
Average&0.612&0.65&0.658&0.689&0.712&\textbf{0.714}\\
\hline
\end{tabular}\label{task1}
\end{table}

In table \ref{task1}, the bold figures show the best values achieved
for corresponding subtasks. We can see that fusion features are more
effective for most subtasks. According to the average values, we
achieve the best result with fusing all the features(LHCT), which
indicates that the features extracted from different views can be
complementary for the DC2010 classification task.

Table \ref{task2} shows the comparison of Web content quality
assessment performance with different scale features for English
task. The features used here is the same as that on Task1.

\begin{table}[!htp]\label{task2}
 \caption{Comparisons of Web content quality assessment performance with
different features on Task2(NDCG)}
 \centering
\begin{tabular}[!htp]{p{1.8cm}p{0.65cm}p{0.65cm}p{0.65cm}p{0.65cm}p{0.65cm}p{0.65cm}}
\hline Task&L&H&C&T&HCT&LHCT\\
\hline \hline

English Task&0.888&0.914&0.918&0.933&\textbf{0.936}&0.935\\
\hline
\end{tabular}\label{task2}
\end{table}

In table \ref{task2}, we can see that link features gives the least
effective result, and TFIDF features show the highest score. The
fused features, such as HCT and LHCT, improve the performance
slightly.

Table \ref{task3} gives the comparison of Web content quality
assessment performance with different scale features for French and
German task. In view of all the labeled hosts are used for the
multilingual quality task, we only employ the page level link
features and host level link features to avoid the semantic
influence of different languages.

\begin{table}[!htp]\label{task3}
 \caption{Comparisons of Web content quality assessment performance with
different features on Task3(NDCG)}
 \centering
\begin{tabular}[!htp]{p{2.1cm}p{1.3cm}p{1.3cm}p{1.3cm}}
\hline Task&L&H&LH\\
\hline \hline
German Task&0.792&\textbf{0.87}&0.854\\
French Task&0.805&\textbf{0.84}&0.833\\
\hline
Average&0.799&\textbf{0.855}&0.844\\
 \hline
\end{tabular}\label{task3}
\end{table}

In table \ref{task3}, we can see that host level link features are
more effective for the cross-linguistic quality tasks.The host level
link features and page level link features we use are not
complementary.

According to the previous description of NDCG performance on all the
tasks, we can see that the host level link features are robust for
most tasks. We can also find that multi-scale features fusion are
necessary for statistical Web content quality assessment.

\section{Conclusions}
In this paper, we explore multi-scale features that may determine
the quality of a host and develop automatic statistical methods to
estimate Web content quality.

The effectiveness of the multi-scale features is analyzed on DC2010
benchmark. The experiments show that the features from different
perspectives have different identification ability and can
complement each other in some degree. For most tasks, we achieve the
best evaluation results with fused features. The experiments also
illustrate the feasibility of the proposed Web content quality
assessment strategy.

Compared with our previous work on Web spam detection\cite{wsc2008},
this paper has the following differences: (a). In the aspect of
targets, \cite{wsc2008} is a detection question, but this paper aims
at a ranking problem. (b). In respect of methods, \cite{wsc2008}
focus on improving the AUC performance of binary classification, but
this paper draws support from the posterior probability of multiple
classification to rank the Web content quality. (c). In terms of
features, we use more features here, for example TFIDF features and
DomainPR related features, etc.

Future work involves: (a). Extract more features, such as natural
language processing features. (b). Explore effective feature fusion
strategy.  (c). Study new quality assessment algorithms, such
as learning the idea of RankBoost\cite{rankboost}.


\section{Acknowledgments}

We would like to thank Hungarian Academy of Sciences, European
Archive Foundation and L3S Hannover for providing the DC2010 data
set, thank DC2010 organization committee for providing the check and
measure script, and thank the reviewers for giving quite a few
suggestions for the paper revision.

This work is supported by by
projects NSFC (No.61005029), the fund for the Doctoral Research of
Shandong Academy of Sciences under Grant No. 2010-12 and the
National Basic Research Program of China under Grant No.
2009CB320502. 

\bibliographystyle{abbrv}

\end{document}